\documentclass[conference]{IEEEtran}
\usepackage[colorlinks]{hyperref}
\IEEEoverridecommandlockouts
\usepackage{subcaption}
\usepackage{cite}
\usepackage{paralist}
\usepackage{multirow}
 \usepackage{siunitx}
\usepackage{amsmath,amssymb,amsfonts}
\usepackage{algorithmic}
\usepackage{graphicx}
\usepackage{textcomp}
\usepackage{xcolor}
\usepackage{pdflscape}
\usepackage{etoolbox} \makeatletter \patchcmd{\@makecaption} {\scshape} {} {} {} \makeatother
\usepackage{rotating} 
\usepackage{adjustbox}
\usepackage{tabularx}
\usepackage{booktabs}
\usepackage{multirow}
\def\BibTeX{{\rm B\kern-.05em{\sc i\kern-.025em b}\kern-.08em
    T\kern-.1667em\lower.7ex\hbox{E}\kern-.125emX}}
\begin{document}

\title{Generalizable and Robust Beam Prediction for 6G Networks: An Deep-Learning Framework with Positioning Feature Fusion}
\author{Yanliang Jin, Yunfan Li, Jiang Jun, Yuan Gao, Shengli Liu, Jianbo Du, Zhaohui Yang, Shugong Xu, \textit{Fellow, IEEE}
\thanks{This work was supported in part by Shanghai Natural Science Foundation under Grant 25ZR1402148, and in part supported by the 6G Science and Technology Innovation and Future Industry Cultivation Special Project of Shanghai Municipal Science and Technology Commission under Grant 24DP1501001. (Yuan Gao is the corresponding authors)} 
\thanks{Yanliang Jin, Yunfan Li, Jiang Jun, Yuan Gao and Shengli Liu are with the School of Communication and Information Engineering, Shanghai University, China, email: jinyanliang@staff.shu.edu.cn, lyf2023@shu.edu.cn, jun\_jiang@shu.edu.cn, gaoyuansie@shu.edu.cn and victoryliu@shu.edu.cn.}
\thanks{Jianbo Du is with the School of Communication and Information Engineering, Xi'an University of Posts and Telecommunications, Xi'an, China, e-mail: dujianboo@163.com.}
\thanks{Zhaohui Yang is with the College of Information Science and Electronic Engineering, Zhejiang University, Hangzhou, China,e-mail: yang\_zhaohui@zju.edu.cn.}
\thanks{Shugong Xu is with the Department of Intelligent Science, Xi’an Jiaotong-Liverpool University, Suzhou, China, email: shugong.xu@xjtlu.edu.cn.}
}

\maketitle
\begin{abstract}
Beamforming (BF) is essential for enhancing system capacity in fifth generation (5G) and beyond wireless networks, yet exhaustive beam training in ultra-massive multiple-input multiple-output (MIMO) systems incurs substantial overhead. To address this challenge, we propose a deep learning–based framework that leverages position-aware features to improve beam prediction accuracy while reducing training costs. The proposed approach uses spatial coordinate labels to supervise a position extraction branch and integrates the resulting representations with beam-domain features through a feature fusion module. A dual-branch RegNet architecture is adopted to jointly learn location-related and communication features for beam prediction. Two fusion strategies, namely adaptive and adversarial fusion, are introduced to enable efficient feature integration. The proposed framework is evaluated on datasets generated by the DeepMIMO simulator across four urban scenarios at 3.5 GHz, following 3GPP specifications, where both reference signal received power (RSRP) and user equipment (UE) location information are available. Simulation results under both in-distribution and out-of-distribution settings demonstrate that the proposed approach consistently outperforms traditional baselines and achieves more accurate and robust beam prediction by effectively incorporating positioning information.

\end{abstract}


\section{Introduction}
With the rapid evolution of 5G and the emergence of sixth generation (6G), wireless networks are undergoing a significant transformation to meet the growing demands for high data rates, low latency, and reliable connectivity \cite{jiang2025towards,gao2019licensed,du2025blockchain,gao2024performance,hu2025MADQN,gao2025stochastic,du2024profit}. Emerging applications such as augmented reality (AR), vehicle-to-everything (V2X), and massive Internet of Things (IoT) further strain existing network capabilities \cite{semiari2019integrated,gao2026csiextra,hu2020fairness,gao2026effective}. To support these requirements, future networks increasingly rely on millimeter-wave (mmWave) communications and massive multiple-input multiple-output (m-MIMO) technologies, which enable wider bandwidths and enhanced spatial resolution through large antenna arrays deployed at base stations (BS) and UE\cite{gao2026nearfield,wang2018spatial,rappaport2017overview,gao2023fair}. By controlling the transmit phase and amplitude of each antenna element, beams can be steered toward desired directions, while larger antenna arrays produce narrower beams that require more accurate control to maintain communication quality. Beam prediction encompasses operations such as beam sweeping, selection, tracking, refinement, and recovery \cite{xue2024survey,brilhante2023literature}, with the objective of establishing and maintaining high-quality directional links between the BS and UE under environmental dynamics, mobility, and blockage. Effective beam prediction directly impacts link reliability, throughput, and user experience, while narrow and highly directional beams further enable spatial reuse and interference mitigation in dense urban deployments \cite{dreifuerst2023massive,gao2023matching}.

Despite its importance, practical beam prediction faces significant challenges. Conventional approaches typically rely on predefined BF codebooks, such as discrete Fourier transform (DFT)–based designs. Although dense codebooks provide high angular resolution, exhaustive beam scanning incurs substantial training overhead, whereas reducing the number of scanned beams lowers overhead at the cost of increased beam misalignment risk and degraded link quality, especially in dynamic urban environments \cite{gapeyenko2017temporal,rangan2014millimeter,martin2018key,hu2024performance}. These issues are further exacerbated in large-scale antenna systems, where conventional algorithms such as compressive sensing or sequential search become computationally intensive and less robust \cite{alkhateeb2014channel,love2008overview}. While some recent studies investigate beam prediction based on full channel state information (CSI), such solutions are often impractical due to high dimensionality, implementation complexity, and feedback latency \cite{alkhateeb2015compressed}. As a result, developing scalable and efficient beam prediction strategies that operate under limited and imperfect information remains a fundamental challenge for future wireless systems \cite{hur2013millimeter}.

In recent years deep learning (DL) has become a powerful tool in wireless communications, achieving remarkable success in a wide range of wireless tasks, including channel estimation, sensing, BF, and resource management~\cite{gao2026dynamic,wang2017deep,3gpp38843,SSnet2025gao,chen20235g,yang20196g,jiang2025mtca,gao2025joint,he2016deep,gao2026UAV,jin2025linformer,gao2025enabling,jin2025context,gao2024c2s,xu2025enhanced,dong2019deep,xia2019deep,li2019deep,lu2025joint,gao2026multiass}. These advances have also motivated the application of DL-based techniques to beam prediction (BP), where models learn spatio-temporal patterns directly from radio-domain observations. Despite their effectiveness, existing DL-based BP methods still face several limitations, including degraded robustness under low signal-to-noise ratio (SNR) conditions, reliance on full CSI, and increased computational complexity~\cite{jalali2024fast,nguyen2022beam,hojatian2021unsupervised,qiao2023intelligent,kim2023joint,dong2019deep,xia2019deep,li2019deep,zhang2023deep}.

To alleviate some of these issues, information-aided designs incorporate auxiliary contextual information, such as position or environmental perception~\cite{va2017position,jiang2022computer,charan2022multi,guo2024predictive,charan2023millimeter}. Accordingly, beam prediction methods that exploit multimodal or auxiliary information have been widely investigated to further enhance performance. From a broader ML perspective, effectively integrating heterogeneous modalities is critical for fully exploiting such auxiliary information, and feature fusion has emerged as an effective technique for performance enhancement. In particular, feature fusion methods have been shown to integrate complementary information and improve generalization across various DL domains~\cite{vielzeuf2018centralnet,sahu2019adaptive,fan2022compound,qin2022domain}.

To enable this integration, we incorporate a feature fusion module inspired by~\cite{sahu2019adaptive} to adaptively combine coordinate-based and beam-domain representations. This design enhances the model’s capability to capture spatial dynamics in complex propagation environments, leading to more robust and reliable beam prediction.

Our main contributions are summarized as follows:
\begin{itemize}

\item \textbf{Systematic review of beam prediction methods:} 
We provide a structured review of recent beam prediction approaches, covering both non-assisted and information-aided methods, together with representative case studies based on 3GPP TR~38.843 (Case~1 and Case~2)~\cite{3gpp38843}. This review highlights key methodological trends and practical limitations in existing studies.

\item \textbf{Position-assisted feature fusion framework:} 
We propose a beam prediction framework that separately extracts beam-domain and location-related features and integrates them via an adaptive fusion mechanism. By leveraging spatial coordinates as supervision during training, the model learns geometry-aware representations that significantly improve prediction accuracy and robustness.

\item \textbf{Comprehensive performance evaluation:} 
Extensive experiments under both in-distribution and out-of-distribution urban scenarios demonstrate that the proposed position-aware framework consistently outperforms baseline models without location assistance, confirming the effectiveness of spatial feature integration.
\end{itemize}

\textit{Notation:}
Bold uppercase letters (e.g., $\mathbf{A}$, $\mathbf{B}$) denote matrices, while bold lowercase letters (e.g., $\mathbf{a}$, $\mathbf{b}$) denote vectors. Lowercase or uppercase letters (e.g., $a$, $A$) denote scalar quantities.  The superscript $(\cdot)^{\mathrm{T}}$ represents the transpose of a matrix or a vector. $\mathbb{C}^{m\times n}$ denotes the complex space of $m \times n$ matrices, and $\mathbb{C}^{m}$ denotes the $m$-dimensional complex vector space.The imaginary unit is represented by $j = \sqrt{-1}$.The notation $\|\cdot\|^2$ denotes the squared Euclidean norm.

\section{Related Work}
This section reviews beam prediction methods, including model-based and learning-based approaches defined in 3GPP~TR~38.843~\cite{3gpp38843}, as well as information-assisted schemes using position or sensing information.

\subsection{Beam Prediction for Cellular Networks}

Beam prediction in cellular networks has evolved in parallel with the development of beam management (BM) mechanisms in 3GPP NR. Early studies focused on exhaustive or hierarchical beam sweeping to establish directional links, and systematically evaluated standard-compliant BM procedures based on synchronization signal blocks (SSBs) and CSI-RS under mobility, blockage, and non-line-of-sight (NLOS) conditions \cite{giordani2018tutorial,fernandes2021beam}. These works revealed that conventional BM performance is highly sensitive to user mobility, angular dispersion, and carrier frequency, leading to substantial measurement overhead and access latency, particularly at mmWave frequencies \cite{fernandes2021beam}.

To mitigate these limitations, recent studies show that machine learning–based beam prediction can reduce measurement overhead while maintaining link reliability \cite{khan2023machine,xu2023performance}.Based on this evolution, 3GPP~TR~38.843~\cite{3gpp38843} formalizes AI/ML-assisted beam prediction into two representative scenarios, namely spatial-domain (Case~1) and temporal-domain (Case~2) beam prediction.
\begin{table*}[htbp]
\centering
\caption{Technical comparison of BM methods.}
\setlength{\tabcolsep}{3.5pt} 
\renewcommand{\arraystretch}{1.32} 

\begin{tabular}{
p{1.2cm}|
p{1.8cm}|
p{2.4cm}|
p{2.3cm}|
p{2.1cm}|
p{3.5cm}|
p{2.9cm}
}
\toprule
\textbf{Ref} &
\textbf{Auxiliary Info} &
\textbf{Input Features} &
\textbf{Method} &
\textbf{Output} &
\textbf{Strengths} &
\textbf{Limitations} \\
\midrule

\cite{3gpp38.214}\cite{3gpp38.802} &
None &
AoD/AoA sampling codebooks &
Hierarchical codebook &
Beam index &
Low complexity; easy to implement on hardware &
Low accuracy; high training overhead \\
\midrule

\cite{khan2024low} &
None &
RSRP &
FCNN &
Beam index &
Low complexity; reduces overhead &
Medium accuracy; poor generalization \\
\midrule

\cite{jalali2024fast} &
None &
RSRP; beam history &
DNN / MLP &
Beam probability &
Fast; reduces overhead &
Medium accuracy; poor generalization \\
\midrule

\cite{nguyen2022beam} &
Orientation (IMU); RSRP &
RSRP; 3D rotation matrix &
LSTM-RNN &
Beam probability &
Good for rotation; improves accuracy &
IMU noise; extra IMU overhead; limited generalization \\
\midrule

\cite{va2017position} &
Position; beam history &
Position; Tx/Rx direction angles &
LtR ranking &
Top-K beams &
Position-assisted; improves performance; reduces overhead &
Requires large dataset; high kernel complexity; requires precise position input \\
\midrule

\cite{cometti2025location} &
Position; online beam power &
Position; beam power; beam index &
GMM + Iterative Probabilistic Search &
Top-K beams &
Online update; reduces overhead; high accuracy &
Requires large dataset; high GMM complexity; requires precise position input \\
\midrule

\cite{lin2024multi} &
CSI; position; camera images &
CSI; position; multi-view images &
Multi-modal DNN (PRSN + BPSN) &
RSU; beam pair &
Visual robustness; high accuracy &
Requires multiple cameras; CSI estimation overhead; requires position input \\
\midrule

\textbf{Our} &
RSRP; position &
RSRP + position (training); RSRP only (inference) &
RegNet + feature fusion &
Beam probability &
High accuracy; reduces overhead; no position needed at inference; leverages spatial information; bridges position-aided and radio-only methods &
Requires large dataset; higher model complexity; requires accurate position labels during training \\
\bottomrule
\end{tabular}
\label{tab:bm_comparison_final}
\end{table*}

\textbf{Case~1:} 
The task is to infer the optimal downlink beam(s) in a target set~A from a smaller set of measured beams~B, reducing measurement overhead while preserving link quality.

\textbf{Case~2:} 
The goal is to predict the optimal beam in a future time slot by exploiting temporal correlations in historical beam measurements, enabling proactive adaptation under mobility or blockage.

For both cases, \cite{3gpp38843} defines standardized key performance indicators (KPIs) for evaluating AI/ML-based BM. 
Among them, Top-1 beam prediction accuracy is the primary metric and is most widely adopted in existing studies, while additional KPIs consider generalization robustness as well as computational complexity and inference latency for practical deployment.

Overall, beam prediction improves cellular systems by reducing measurement overhead, enhancing link robustness, and enabling more efficient and reliable communication, making it a key enabler for intelligent 5G-Advanced and beyond networks.

\subsection{Beam Prediction Algorithms}

This subsection reviews beam prediction algorithms, which can be broadly categorized into model-based methods relying on analytical and geometry-driven formulations, and AI-based methods that learn beam mappings directly from communication signals.

\subsubsection{Model-Based Beam Prediction Methods}
Model-based beam prediction relies on analytical and geometry-based signal models. 
These methods can be categorized into Case 1 and Case 2.

Case 1 aims to identify the optimal transmit–receive beam pair from a limited set of measurements, avoiding exhaustive beam sweeping. In 5G NR, this process is commonly realized through hierarchical BM based on resolution-refined codebooks, employing coarse-to-fine beam sweeping and CSI-RS-based refinement~\cite{3gpp38.214,3gpp38.802}. Beyond standardized schemes, model-based beam prediction methods address diverse deployment scenarios and optimization objectives through different modeling assumptions, including hierarchical and adaptive search, geometry-driven feature construction, and interference-aware formulations~\cite{xiao2016hierarchical,marandi2018adaptive,yoon2021off,jiao2024joint,ganji2024terra,rezaei2025enhancing,wang2025beam}.

Case 2 exploits motion dynamics, temporal correlations, or auxiliary sensing to anticipate future optimal beams and reduce frequent retraining. 
Temporal beam prediction methods have evolved from localization and multi-sensor fusion~\cite{shimizu2018millimeter}, to filtering-based tracking using particle or Kalman filters~\cite{chung2020adaptive,ali2021orientation,matsuno2023high}, and further to predictive and cooperative designs leveraging vehicle-state prediction and cross-band sensing~\cite{guo2024predictive,chen2024cooperative,sun2025fusion}.

In summary, model-based beam prediction methods defined in 3GPP~TR~38.843 provide interpretable baselines but suffer from limited robustness and scalability due to simplified modeling assumptions. AI/ML-based approaches overcome these limitations through data-driven learning~\cite{khan2023machine}.

\subsubsection{Deep-Learning-Based Beam Prediction Methods}

Non-aided deep-learning-based beam prediction learns beam mappings directly from radio-domain inputs and follows the Case~1 and Case~2 categorization in 3GPP~TR~38.843.

Case~1
\textit{(i) RSRP-based inputs.}
RSRP-based beam prediction methods combine the sparsity of RSRP measurements with machine learning models for non-aided beam inference~\cite{hojatian2021unsupervised,zuo2022artificial,xu2023performance,jalali2024fast,khan2024low,khan2024generalization}. 
Recent works further focus on jointly optimizing probing efficiency and prediction accuracy under challenging propagation conditions~\cite{zhang2025kpi,xue2025integrated}.

\textit{(ii) CSI-based inputs.}
CSI-based deep-learning methods explore the relationship between CSI and beam selection by learning data-driven mappings from CSI observations to beam indices or codebooks~\cite{zhang2020learning,ma2021deep,chen2023active}. 
More recent studies further incorporate reinforcement learning to adapt beam decisions from time-varying CSI, improving robustness under user mobility~\cite{qiao2023intelligent}.

\textit{(iii) Other radio-domain inputs.}
Beyond RSRP and CSI, existing works exploit alternative radio-domain measurements for beam prediction by learning data-driven mappings from pilot responses, received-power patterns, or raw antenna signals~\cite{echigo2021deep,cousik2022deep,ktari2024neural,jamshidi2024practical}. 
Recent studies further adopt reinforcement learning to optimize beam selection and link adaptation under dynamic channel conditions~\cite{kim2023joint,zhang2023deep}.

Case~2
\textit{(i) RSRP-based inputs.}
RSRP-based temporal beam prediction models beam dynamics under mobility using sequential learning on historical RSRP observations~\cite{si2020deep,domae2022machine}. 
Subsequent studies enhance this paradigm by combining spatial feature extraction with advanced temporal modeling, including CNN-, FC-, LSTM-, and Transformer-based architectures as well as UE-side inference, to improve multi-step prediction accuracy and robustness under high mobility~\cite{li2023machine,wang2023deep,bai2023ai,li2024artificial,xu2024parallel,xu2023performance}.

\textit{(ii) CSI-based inputs.}
CSI-based temporal beam prediction exploits historical CSI using sequential learning to model beam dynamics under mobility~\cite{guo2019machine}. This paradigm is further enhanced by attention and hybrid learning filtering methods~\cite{wang2022attention,zhang2024hybrid}.

\textit{(iii) Other radio-domain inputs.}
Beyond CSI and RSRP, existing works exploit alternative radio-domain representations for temporal beam prediction, including beam image features, compressive measurements, and near field channel observations. 
These methods typically combine deep learning or reinforcement learning with temporal modeling to enable efficient beam tracking under mobility~\cite{jin2023beam,wang2024spatial,wang2025near}.

In summary, non-aided AI-based beam prediction reduces sweeping overhead and improves accuracy through data-driven learning, but remains sensitive to data quality and limited in interpretability.

\subsection{Additional Information-Assisted Beam Prediction}

Beam prediction can be enhanced by exploiting auxiliary information such as position, sensing data, or cross-frequency CSI, which provides richer spatial temporal context and improves robustness under mobility and blockage~\cite{jin2024efficient,gao2026sidelink}. 
This benefit comes at the cost of additional signaling and implementation complexity.

\subsubsection{Position- and Angle-Aided Approaches}

Position- and angle-aided beam prediction exploits geometric priors to accelerate beam selection and improve robustness under mobility. 
Early studies primarily relied on explicit location or angular mappings, learning associations between user position, road geometry, or multipath fingerprints and optimal beams, with online refinement to mitigate localization errors~\cite{va2017position,va2017inverse,va2019online}. 
Subsequent works extended this paradigm by integrating position, orientation, motion state, and environmental context into machine-learning or reinforcement-learning frameworks for more stable beam selection in dynamic scenarios~\cite{wang2018mmwave,wang2019mmwave,rezaie2020location,wang2019reinforcement,xu2021data,nguyen2022beam}.

More recent approaches increasingly incorporate richer temporal and probabilistic modeling. 
These include generative and probabilistic formulations that jointly infer position and beam likelihoods~\cite{cometti2025location}, as well as sequence-learning methods that exploit historical position, motion, RSRP, or beam sequences using RNN-, LSTM-, or attention-based architectures to generalize across mobility regimes~\cite{kutty2023deep,bai2023ai,liu2025sa}. 
Lightweight learning models combining inertial sensing or orientation cues have also been explored for efficient angle tracking~\cite{chung2024effective}, alongside predictive and multi-cell coordination frameworks that leverage trajectory-aware modeling for beam scheduling~\cite{guo2024predictive}.

In parallel, hybrid learning and model-based strategies combine geometric tracking with data-driven prediction to improve search efficiency and service continuity~\cite{wang2023improved,wang2024dual}. 
Finally, cooperative and multi-sensor approaches exploit spatial diversity across multiple base stations or sensing modalities, such as multi-camera or multi-RSU observations, to enhance beam inference in dense and urban environments~\cite{alkhateeb2018deep,lin2024multi}.

\subsubsection{Vision- or Environment-Assisted Approaches}

Vision- and environment-assisted beam prediction exploits explicit geometric and scene awareness to complement radio-domain features, enabling blockage detection, LoS identification, and reflection-path inference. 
Early studies demonstrated that visual or 3D sensing modalities, such as RGB cameras and LiDAR, can be directly mapped to beam indices using deep learning, establishing feedback-free or sensing-assisted beam alignment frameworks~\cite{alrabeiah2020millimeter,marasinghe2022lidar,jiang2022computer,ahn2022toward,charan2022multi,jiang2022lidar,charan2023millimeter}. 
Related works also explored latent-space representations, such as channel charting, to enable lightweight beam tracking from reduced-dimensional channel features~\cite{kazemi2022channel,zhang2023beam}.

More recent approaches focus on improving robustness and scalability by incorporating advanced learning paradigms. 
These include ranking-based and curriculum-learning strategies for vision-based beam selection~\cite{charan2023camera,wang2024vision}, Transformer-based object detection and scene understanding for more accurate angle estimation~\cite{son2024object}, and multi-modal fusion of vision, LiDAR, radar, and GPS information using deep or attention-based architectures~\cite{chen2023beam,yanpeng2024sensing,ashour2024perception,wang2024intelligent,yeo2024multi,ghassemi2024multi}. 
In parallel, radar-assisted and environment-aware designs integrate sensing and communication through filtering, reinforcement learning, or federated learning to enhance adaptability in dynamic and distributed deployments~\cite{wang2024mmwave,zhou2024radar}.

\subsubsection{Cross-Domain Learning for Beam Prediction and Coordination}

Cross-domain learning exploits correlations across frequency bands, modalities, or network domains to reduce beam-training overhead and enable scalable coordination. 
Early studies demonstrated that sub-6~GHz channels can be mapped to mmWave beam directions using deep learning, establishing cross-band prediction frameworks that avoid exhaustive in-band training~\cite{sim2020deep,alrabeiah2020deep}. 
Subsequent works extended this idea by addressing temporal misalignment and jointly learning heterogeneous-band features through recurrent or fusion-based neural networks, enabling more robust cross-frequency beam prediction~\cite{ma2021deep,ma2022deep,sheng2023two}.

More recent studies have expanded cross-domain learning to multi-user, higher-frequency (e.g., FR3), and coordinated scenarios. 
Representative approaches combine deep learning with sequential modeling, graph-based coordination, or beamspace representations to support proactive BM, hybrid BF, and spectral-transfer learning across bands~\cite{vankayala2023efficient,deng2025csi,chen2025frequency}. 
In parallel, cross-domain beam prediction has been formulated as a multi-agent or contextual decision-making problem, leveraging surrogate models, graph neural networks, and attention-based learning to enable coordinated beam selection under partial observations~\cite{li2025joint,li2025contextual}.

Overall, auxiliary-information-assisted beam prediction enhances spatial awareness beyond radio-only inputs by leveraging geometric priors, visual and environmental perception, and cross-frequency knowledge transfer. 
By exploiting position, sensing, and multi-band contextual information, these approaches improve robustness under mobility and blockage while reducing training and pilot overhead. 
Such perception- and context-driven strategies provide a scalable and reliable foundation for low-latency, intelligent BM in future 6G and ISAC-integrated systems.

To provide a concise overview of auxiliary-information-assisted beam prediction, Table~\ref{tab:bm_comparison_final} summarizes representative frameworks in terms of auxiliary inputs, model architectures, prediction targets, and practical strengths and limitations. While many existing works incorporate positional information, they typically rely on explicit position inputs at inference time or adopt loosely coupled multi-task or multi-branch designs, and often require highly accurate localization, dedicated sensing modalities, or online probing. 

In contrast, this study focuses on exploiting positional semantics only during training, where a feature fusion module is introduced to effectively integrate positional and beam-domain representations and mitigate distribution mismatch between heterogeneous features. 
As a result, positional information is not required at inference time, and beam prediction is performed solely based on RSRP measurements, reducing deployment overhead while preserving the spatial benefits of position-aided learning.

\section{System Model}
We consider MIMO system comprising a single BS and multiple mobile UEs. The BS is equipped with a large-scale antenna array, while each UE is equipped with a smaller antenna array. The objective is to predict the optimal subset of downlink beam directions for efficient beam scanning.
\subsection{Channel Model}
In this work, we adopt the geometric channel model provided by the DeepMIMO dataset, which is based on ray-tracing simulations. In our simulation, each BS is divided into three sectorized coverage regions, and the pair $(b, u)$ represents the specific sector $b$ and the user $u$ associated with it. Each sector employs its own directional antenna array, consistent with the sectorized deployment model described in 3GPP TR 38.843~\cite{3gpp38843}. For each BS-UE pair $(b, u)$, the downlink channel vector on the $k$-th orthogonal frequency division multiplexing (OFDM) subcarrier, denoted by $\mathbf{h}_{k}^{b,u}$, is modeled as a summation over $L$ propagation paths. Each path $\ell$ is characterized by its received power $\rho_\ell$, path delay $\tau_\ell^{b,u}$, and phase $\vartheta_\ell^{b,u}$, as well as its azimuth and elevation angles of departure $(\phi_{\text{az}}^{b,u}, \phi_{\text{el}}^{b,u})$. The BS is equipped with a three dimensional (3D) uniform planar array (UPA), and the array response vector $\mathbf{a}(\phi_{\text{az}}^{b,u}, \phi_{\text{el}}^{b,u})$ is constructed based on the departure angles of each path. Let $B$ denote the channel bandwidth and $K$ be the total number of OFDM subcarriers. Then, the downlink channel is given by\cite{alkhateeb2019deepmimo}:
\begin{equation}
\mathbf{H}^{b,u} = \sum_{\ell=1}^{L} \sqrt{\frac{\rho_\ell}{K}} 
\exp\left( j\left( \vartheta_\ell^{b,u} + \frac{2\pi k}{K} \tau_\ell^{b,u} B \right) \right)
\mathbf{a}\left( \phi_{\text{az}}^{b,u}, \phi_{\text{el}}^{b,u} \right).
\label{eq:deepmimo_channel}
\end{equation}

To capture the directional characteristics of the transmitted signal, the array response vector of the BS is constructed based on the angles of departure. Specifically, for each path $\ell$ corresponding to the $(b,u)$ transmitter–receiver pair, the array response vector $\mathbf{a}(\phi_{\text{az}}^{b,u}, \phi_{\text{el}}^{b,u})$ is computed as the Kronecker product of the steering vectors along the $z$, $y$, and $x$ spatial axes, respectively. That is,
\begin{equation}
\mathbf{a}(\phi_{\text{az}}^{b,u}, \phi_{\text{el}}^{b,u}) = 
\mathbf{a}_z(\phi_{\text{el}}^{b,u}) \otimes 
\mathbf{a}_y(\phi_{\text{az}}^{b,u}, \phi_{\text{el}}^{b,u}) \otimes 
\mathbf{a}_x(\phi_{\text{az}}^{b,u}, \phi_{\text{el}}^{b,u}).
\label{eq:array_vector}
\end{equation}

Here, the sub-vectors $\mathbf{a}_x(\cdot)$, $\mathbf{a}_y(\cdot)$, and $\mathbf{a}_z(\cdot)$ denote the 1D array response vectors along the $x$-, $y$-, and $z$-axis, respectively, which are given by:
\begin{align}
\mathbf{a}_x(\phi_{\text{az}}^{b,u}, \phi_{\text{el}}^{b,u}) &= 
\left[
1, e^{jkd \sin(\phi_{\text{el}}^{b,u}) \cos(\phi_{\text{az}}^{b,u})}, \dots, \right. \nonumber \\
&\quad \left. e^{jkd(M_x - 1) \sin(\phi_{\text{el}}^{b,u}) \cos(\phi_{\text{az}}^{b,u})}
\right]^{\mathrm{T}}, \label{eq:ax} \\[1pt]
\mathbf{a}_y(\phi_{\text{az}}^{b,u}, \phi_{\text{el}}^{b,u}) &= 
\left[
1, e^{jkd \sin(\phi_{\text{el}}^{b,u}) \sin(\phi_{\text{az}}^{b,u})}, \dots, \right. \nonumber \\
&\quad \left. e^{jkd(M_y - 1) \sin(\phi_{\text{el}}^{b,u}) \sin(\phi_{\text{az}}^{b,u})}
\right]^{\mathrm{T}}, \label{eq:ay} \\[1pt]
\mathbf{a}_z(\phi_{\text{el}}^{b,u}) &= 
\left[
1, e^{jkd \cos(\phi_{\text{el}}^{b,u})}, \dots, 
e^{jkd(M_z - 1) \cos(\phi_{\text{el}}^{b,u})}
\right]^{\mathrm{T}}. \label{eq:az}
\end{align}
\subsection{Codebooks}
BF is a key technique that enables directional signal transmission and reception by exploiting the spatial characteristics of antenna arrays. 
The fundamental principle of BF is to adjust the initial phase and amplitude of the signals transmitted by each antenna element, 
such that the electromagnetic waves emitted from different antennas constructively interfere in a desired direction and destructively interfere in other directions. 
Through this process, a strong beam is formed toward the intended angle, thereby enhancing the signal strength and suppressing interference from undesired directions.
To support BF in the system, we adopt predefined codebooks to steer the beam in specific directions. The codebooks are designed based on DFT principles and vary in angular resolution depending on the configuration. 

Each codebook consists of a finite set of beam directions generated by quantizing the azimuth and elevation domains. We generate DFT codebooks with sizes $X_1 \times X_2$ .The codebooks are critical for determining the candidate beam directions during beam scanning.

Codebook defines a set of beam directions in the 2D angular domain $[\varphi, \theta]$, where
\begin{equation}
\theta \in \mathbf{H}_{X_1} = \left[\theta_{\min} + \frac{120}{X_1} \cdot k\right], \quad k = 0, \dots, X_1{-}1,
\end{equation}
\begin{equation}
\varphi \in \mathbf{V}_{X_2} = \left[-90 + \frac{180}{X_2} \cdot l\right], \quad l = 0, \dots, X_2{-}1,
\end{equation}
where $\theta_{\min}$ is the starting azimuth angle of each sector, and $X_1 \times X_2$ represents the angular resolution of the codebook, where $X_1$ denotes the resolution in the azimuth (horizontal) domain and $X_2$ denotes the resolution in the elevation (vertical) domain.

Overall, each codebook corresponds to a spatial combination of beam patterns pointing toward different angular directions. 
The codebook matrix is denoted by $\mathbf{B} \in \mathbb{C}^{(X_1 \times X_2) \times (M \times N)}$, 
where $X_1 \times X_2$ represents the total number of beams and $M \times N$ denotes the total number of antenna elements at the BS.

Each row of the codebook matrix $\mathbf{B}$~\cite{huang2020dft,suh2016construction} represents a BF vector. 
We use $\mathbf{f}$ to denote the BF vector form and $\mathbf{F}$ to denote the BF matrix form. 
Assuming an $8 \times 8$ UPA, the corresponding BF matrix for all antenna elements is denoted as $\mathbf{F}$. 
When $\mathbf{F}$ is unfolded by rows or columns, it forms the BF vector $\mathbf{f}$ used to steer the beam in a specific direction.

\begin{equation}
\label{bf_v}
\textbf{F}=  
\begin{bmatrix}  
  a^{7} & a^{7}b & \cdots  & a^{7}b^{7}  \\  
  a^{6} & a^{6}b & \cdots  & a^{6}b^{7} \\  
  \vdots & \vdots & \ddots & \vdots \\  
  1 & b & \cdots  & b^{7}  
\end{bmatrix}
\end{equation}where:

\begin{equation}
\label{}
a = e^{j\pi\sin\theta},
\end{equation}and
\begin{equation}
\label{}
b = e^{-j\pi\cos\theta\cos\varphi}.
\end{equation}

\subsection{Beam Prediction}

We model the received signal after BF as
\begin{equation}
    \mathbf{y} = \mathbf{H} \mathbf{f} \mathbf{x} + \mathbf{n},
\end{equation}
where $\mathbf{H} \in \mathbb{C}^{N \times M}$ denotes the downlink channel matrix, 
with $M$ representing the number of transmit antennas at the BS and $N$ the number of receive antennas at the UE. 
$\mathbf{f} \in \mathbb{C}^{M \times 1}$ is the transmit BF vector, 
$\mathbf{x} \in \mathbb{C}^{1 \times 1}$ (or scalar symbol) represents the transmitted data symbol, 
and $\mathbf{n} \in \mathbb{C}^{N \times 1}$ denotes the additive noise vector, modeled as a circularly symmetric complex Gaussian random vector with $\mathbf{n} \sim \mathcal{CN}(\mathbf{0}, \mathbf{I})$. 
Accordingly, the received signal $\mathbf{y} \in \mathbb{C}^{N \times 1}$ is a complex vector observed at the UE.

In practical systems, beam prediction is performed by selecting the best BF vector from a predefined codebook $\mathcal{B} = \{\mathbf{f}_1, \mathbf{f}_2, \dots, \mathbf{f}_K\}$, where each $\mathbf{f}_i$ represents a candidate beam and $I$ is the total number of beams in the codebook. The optimal beam index $i^\star$ is selected by maximizing the received power metric, commonly referred to as the RSRP, which is defined as
\begin{equation}
\text{RSRP}_i = \|\mathbf{H} \mathbf{f}_i\|^2.
\label{eq:rsrp}
\end{equation}

\begin{figure*}[t]
    \centering
    \includegraphics[width=0.95\textwidth]{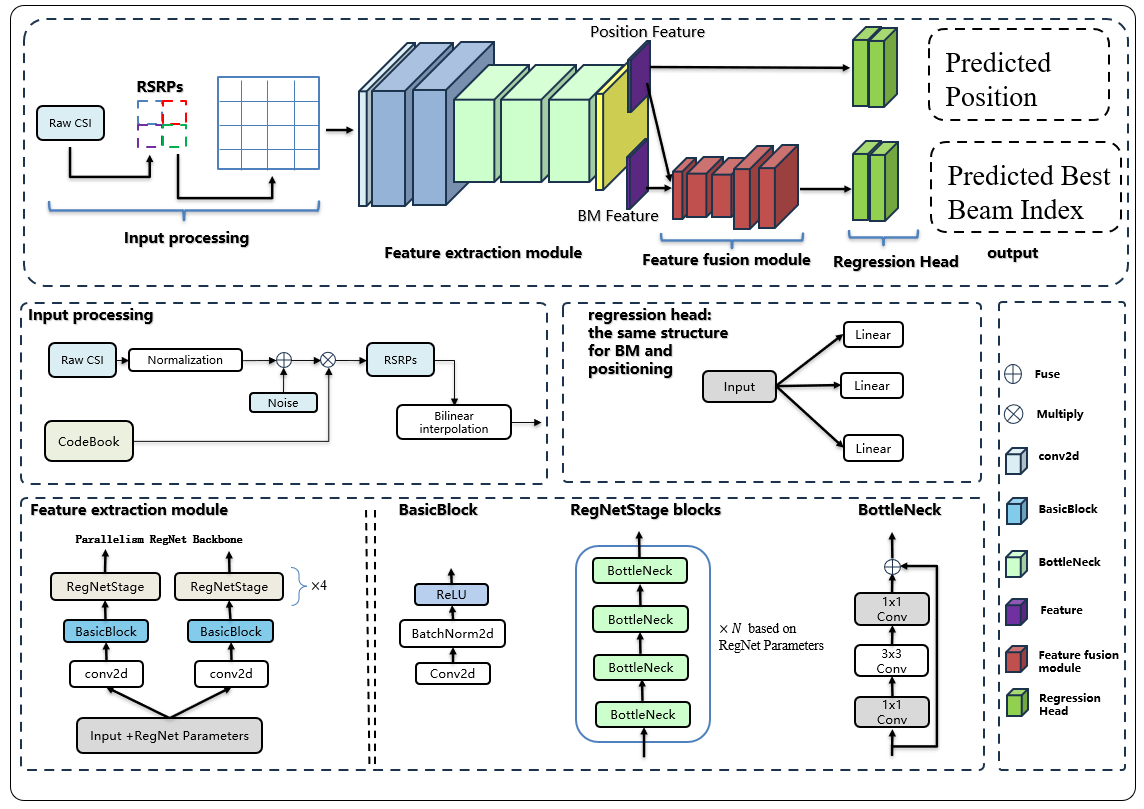}
    \caption{Overall architecture of the proposed model. The pipeline begins with RSRP-based beam inputs derived from different codebooks. A dual-branch backbone extracts beam-specific and position-specific features. These are integrated via a feature fusion module and ultimately used to predict both beam prediction and positioning outputs across multiple regions.}
    \label{fig:model_structure}
\end{figure*}

\begin{figure}[htbp]
    \centering
    \includegraphics[width=1\linewidth]{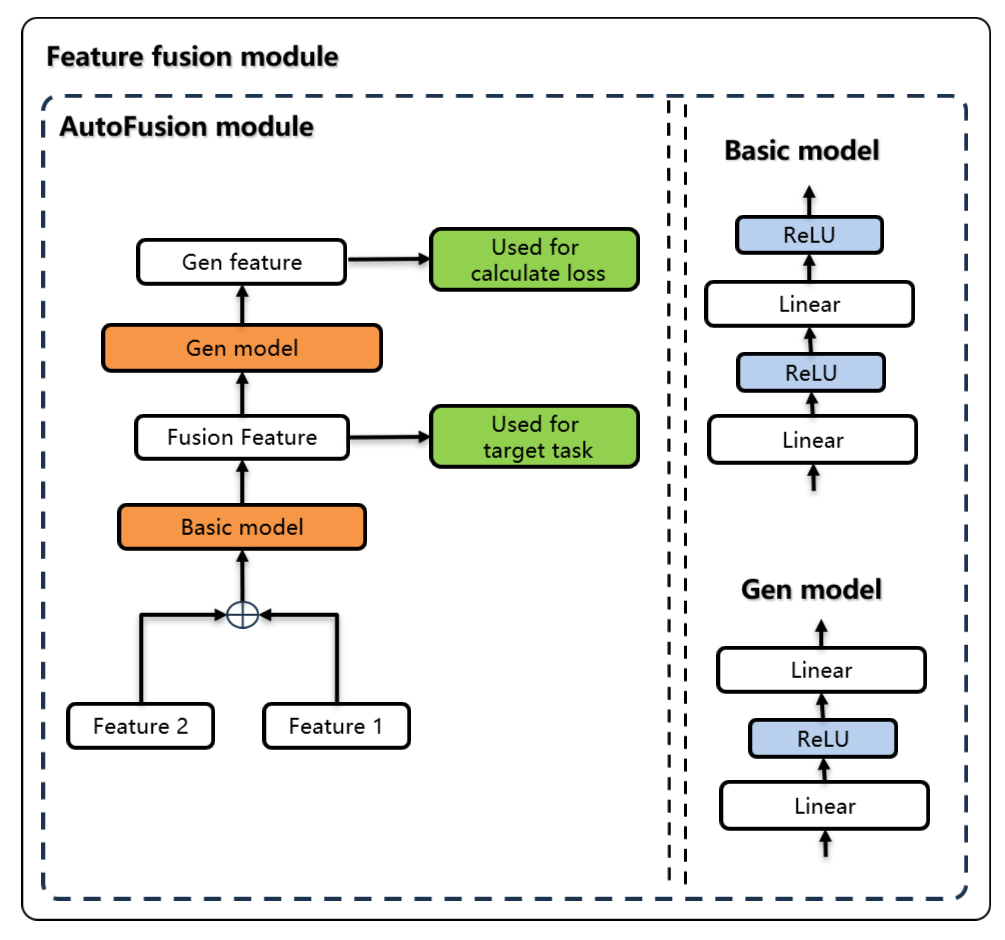}
    \caption{Structure of the AutoFusion module, adapted based on the original design in the reference. The module learns to combine two feature branches and generate a fused representation suitable for both loss supervision and target prediction.}
    \label{fig:autofusion}
\end{figure}

The beam prediction process then becomes a search problem:
\begin{equation}
i^\star = \arg\max_{i \in \{1, \dots, I\}} \text{RSRP}_i.
\end{equation}

In our simulation setup, noise is added directly to the channel vector $\mathbf{H}$ before BF, in order to simulate realistic channel impairments under various SNR conditions. The noise power is denoted as $\mathfrak{N}$, and the received SNR is defined as
\begin{equation}
    \text{SNR} = 10 \log_{10} \left( \frac{\|\mathbf{H}\|^2}{\mathfrak{N}} \right).
\label{eq:snr}
\end{equation}
Given a target SNR value in decibels, denoted by $\gamma_{\text{dB}}$, the corresponding linear-scale noise power can be computed as
\begin{equation}
\mathfrak{N} = \frac{\|\mathbf{H}\|^2}{10^{\gamma_{\text{dB}}/10}}.
\end{equation}
The noise vector is then sampled from a complex Gaussian distribution $\mathcal{CN}(0, 1)$ and added element-wise to $\mathbf{H}$. This method allows us to control the degradation level of the channel by adjusting the SNR, and it enables consistent evaluation of beam prediction performance under varying channel quality.

\section{Proposed AI model}
In this section, we proposed a DL-based model to effectively utilize the predicted position information to enhance beam prediction performance. The proposed model compromises pre-processing, feature extraction, feature fusion and regression head.

\begin{figure}[htbp]
    \centering
    \includegraphics[width=1\linewidth]{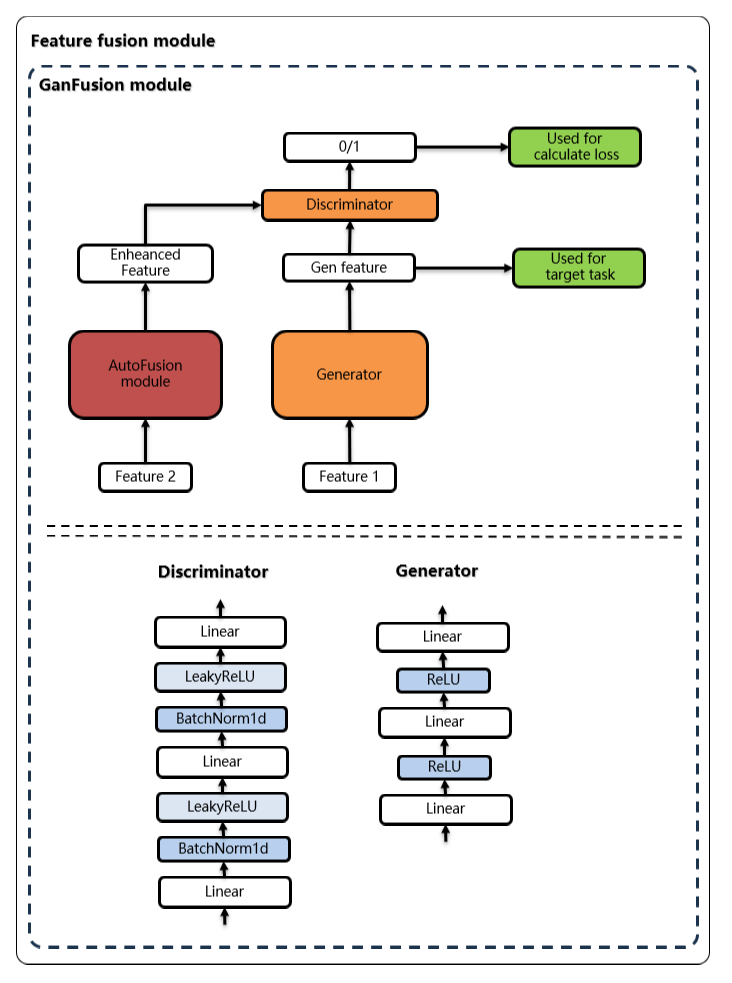}
    \caption{Structure of the GanFusion module. The generator learns to synthesize fused features from one domain, while the discriminator distinguishes them from features generated by the AutoFusion module. The adversarial training encourages the generator to produce task-consistent and domain-aligned representations.}
    \label{fig:ganfusion}
\end{figure}

The architecture of our proposed model is illustrated in Fig.~\ref{fig:model_structure}, where beam-domain and position-domain features are extracted separately and fused via a dedicated module to improve beam prediction performance.

\subsubsection{Pre-processing}
Firstly, the CSI signals are normalized. Before computing the RSRP, noise is added to the normalized CSI based on the noise power calculated using equation (~\ref{eq:snr}). Then, a matrix multiplication between the noise-added CSI and the codebook is performed, followed by taking the absolute square to compute the RSRP as shown in equation (~\ref{eq:rsrp}). Since convolution-based networks are highly sensitive to the scale of input data, and the input data in our task is extremely sparse, we employ bilinear interpolation to expand the data dimensions. For example, a $4 \times 4$ input is linearly mapped to a $64 \times 64$ resolution. This upsampling strategy allows the model to extract features more effectively by increasing the spatial resolution of the input.
Bilinear interpolation is computed using the following formula:
\begin{equation}
\begin{aligned}
f(x, y) = &\; (1 - (x - x_0))(1 - (y - y_0))\, Q_{11} \\
         &+ (x - x_0)(1 - (y - y_0))\, Q_{21} \\
         &+ (1 - (x - x_0))(y - y_0)\, Q_{12} \\
         &+ (x - x_0)(y - y_0)\, Q_{22},
\end{aligned}
\end{equation}where $(x, y)$ denotes the target coordinate, and $Q_{11}, Q_{21}, Q_{12}, Q_{22}$ are the values of the four neighboring points surrounding $(x, y)$ in the source grid.

\subsubsection{Feature extraction}
Our framework employs two parallel RegNet-based\cite{radosavovic2020designing} backbones to extract task-specific features from the upsampled RSRP data. One branch focuses on spatial representations for the positioning task, while the other learns directional features for beam prediction. We adopt a RegNet structure similar to RegNet-50, with specific parameters computed from $(w_a = 31.41, w_0 = 96, w_m = 2.24, d = 22)$, ensuring a balance between representational power and computational efficiency. The corresponding RegNet widths are generated using the linear rule 
$\bar{w}_i = w_0 + w_a i$ and the quantization 
$w_i = w_0\, w_m^{\left\lfloor \frac{\log(\bar{w}_i / w_0)}{\log(w_m)} \right\rceil}$, where $w_i$ denotes the quantized channel width of the $i$-th block in the RegNet backbone. Each branch processes the same input independently, enabling the network to capture distinct feature modalities optimized for their respective objectives.

As illustrated in Fig.~\ref{fig:model_structure}, each branch begins with a BasicBlock composed of a $3\times3$ convolution layer, batch normalization, and ReLU activation, which performs initial feature extraction and spatial downsampling. This is followed by four RegNet stages, each consisting of multiple Bottleneck blocks. A Bottleneck block includes a $1\times1$ convolution for channel reduction, a $3\times3$ convolution for feature transformation, and a final $1\times1$ convolution for channel expansion, along with a residual connection. The number of Bottleneck blocks in each stage is determined by the RegNet parameters $(w_a, w_0, w_m, d)$, ensuring scalability and consistent stage widths. This modular structure enables deep hierarchical feature learning with efficient parameter usage.

\subsubsection{Feature fusion}
To enhance the performance of beam prediction, we incorporate position-related features as auxiliary guidance during beam prediction. After independently extracting features from the positioning and beam-related RegNet branches, the fusion module integrates these two representations to enrich the spatial understanding of the beam prediction process. The position-aware features provide valuable geometric context, helping the model better capture its correlation between UE location and optimal beam direction. This fused representation is then used to support the final beam prediction decision.

To effectively integrate the positional and beam-related features extracted by the dual-branch RegNet backbone, we adopt a feature fusion strategy based on AutoFusion and GanFusion.

\subsubsection{AutoFusion module} As shown in Fig. ~\ref{fig:autofusion}, the AutoFusion module is designed to transform concatenated features from two different domains into a more structured and task-relevant representation. It first concatenates the two feature vectors as concatenated ground-truth feature and passes it through the Basic model to generate a fused feature that is better suited for the downstream task. During inference, this fused feature is directly used for prediction.

During training, the fused feature is further processed by the Gen model to produce a Gen feature, which is then compared with the original concatenated input using mean square error loss. This reconstruction loss encourages the fusion process to retain essential information from the original features while reshaping their distribution to better align with the target network. The loss function is defined as follows:

\begin{equation}
J_{Auto} = \left\| \mathbf{f}_{\text{gen}} - \mathbf{f}_{\text{cat}} \right\|^2,
\label{eq:tr_loss}
\end{equation}where $\mathbf{f}_{\text{gen}}$ denotes the generated feature and $\mathbf{f}_{\text{cat}}$ represents the concatenated ground-truth feature.

As illustrated on the right side of Figure~\ref{fig:autofusion}, both the \textit{Basic model} and the \textit{Gen model} are implemented as multi-layer perceptrons (MLPs) with specific architectures. The Basic model consists of two linear layers followed by ReLU activations, while the Gen model uses two linear layers with a single ReLU activation in between. In terms of dimensionality, both models take a 1079-dimensional concatenated feature as input, compress it into a 539-dimensional latent representation, and then reconstruct it back to a 1079-dimensional output.

\subsubsection{GanFusion module} As shown in Fig. ~\ref{fig:ganfusion}, building on top of AutoFusion, the GanFusion module adopts an adversarial training paradigm to further refine the fused feature representations.

In our framework, we first apply the AutoFusion module to enhance the auxiliary feature (Feature 2). We observe that even a single input feature processed by AutoFusion can effectively improve feature expressiveness and enhance model performance. The output is referred to as the Enhanced Feature.
Next, the Generator takes another feature (Feature 1) as input and produces a Gen feature, which is intended to be aligned with the enhanced auxiliary feature. Simultaneously, a Discriminator is trained to differentiate between the Gen feature and the Enhanced Feature. Through this adversarial process, the generator is incentivized to produce fused features that are not only indistinguishable from real enhanced features but also more relevant to the target task.

This adversarial interplay helps align the distribution of the generated features with that of the real features, thereby improving the quality, consistency, and task-relevance of the learned feature representations.

The adversarial loss is defined as:

\begin{align}
\min_G \max_D J_{\text{adv}}(D, G) =\ & 
\mathbb{E}_{\mathbf{f}_{\text{enh}} \sim p(\mathbf{f}_{\text{enh}})}[\log D(\mathbf{f}_{\text{enh}})] \notag \\
& +\ \mathbb{E}_{\mathbf{f}_{\text{gen}} \sim p(\mathbf{f}_{\text{gen}})}[\log (1 - D(\mathbf{f}_{\text{gen}}))],
\label{eq:adv_loss}
\end{align}where $\mathbf{f}_{\text{enh}}$ denotes the real enhanced feature from the AutoFusion module, and $\mathbf{f}_{\text{gen}}$ is the feature generated by the Generator $G$. The discriminator $D(\cdot)$ outputs a probability indicating whether the input is real or fake. The expression $\mathbf{f}_{\text{enh}} \sim p(\mathbf{f}_{\text{enh}})$ means that the enhanced features are sampled from the real distribution learned by the AutoFusion module, and the term $\mathbb{E}_{\mathbf{f}_{\text{enh}} \sim p(\mathbf{f}_{\text{enh}})}[\log D(\mathbf{f}_{\text{enh}})]$ represents the expected log-likelihood that the discriminator assigns to real features. Similarly, the generator $G$ aims to produce $\mathbf{f}_{\text{gen}}$ that fools the discriminator. In addition to adversarial training, the generated feature $\mathbf{f}_{\text{gen}}$ is also used for the target task, such as beam prediction.

As illustrated in the lower half of Fig.~\ref{fig:ganfusion}, both the \textit{Generator} and the \textit{Discriminator} are implemented using MLPs with specific architectural designs. The Generator consists of three linear layers with two ReLU activation functions applied between them. In terms of dimensionality, the Generator takes a 1079-dimensional feature as input, expands it to a 1079×2-dimensional hidden representation, and then outputs a 1079-dimensional generated feature.
The Discriminator, on the other hand, includes three linear layers, interleaved with LeakyReLU activations and BatchNorm1d normalization layers after the first and second hidden layers, enabling stable adversarial training. Its MLP receives a 1079-dimensional feature, processes it through a 1079-dimensional hidden layer, and finally produces a single scalar output for real/fake discrimination.

In our framework, the AutoFusion module and the GanFusion module serve as two distinct feature fusion mechanisms. Although both modules can be independently integrated into the proposed model for multimodal feature aggregation, they are designed based on fundamentally different principles. It is worth noting that the GanFusion module incorporates a component that is conceptually related to AutoFusion; however, this does not alter the fact that they remain two separate fusion modules. Therefore, in the subsequent experiments, we evaluate the performance of each module individually to provide a fair and comprehensive comparison of their respective fusion capabilities.
\subsubsection{Regression head}
To handle different sectors, we employ three separate classification heads, each responsible for selecting the optimal beam within its corresponding sector. This design enables the model to effectively learn spatial features embedded in the RSRP measurements across different sectors. Each classification head outputs a probability distribution over the high-resolution codebook, from which the beam with the highest probability is ultimately selected as the final output.

\begin{figure*}[htbp]
    \centering

    \begin{subfigure}[t]{0.45\textwidth}
        \centering
        \includegraphics[width=\linewidth]{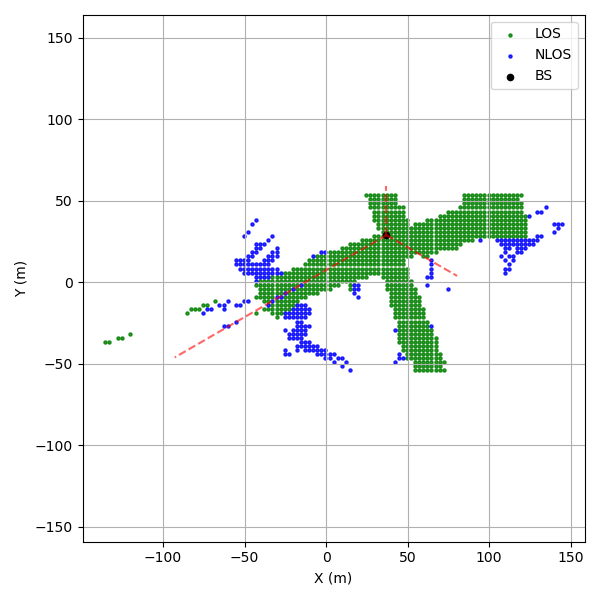}
        \caption{New York}
    \end{subfigure}
    \hfill
    \begin{subfigure}[t]{0.45\textwidth}
        \centering
        \includegraphics[width=\linewidth]{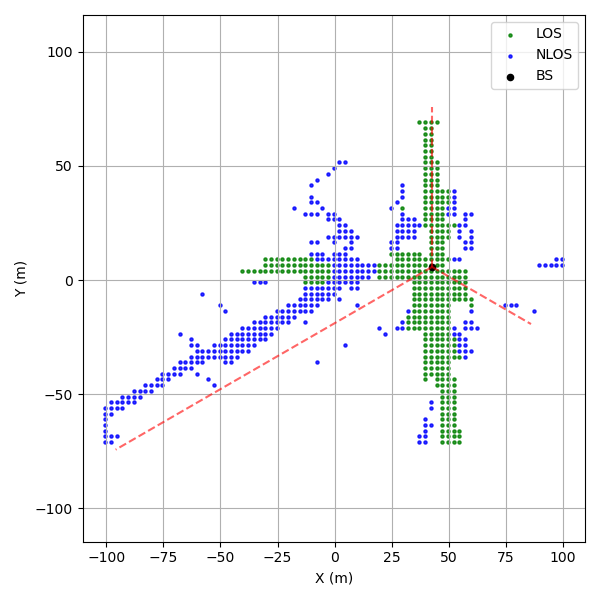}
        \caption{Los Angeles}
    \end{subfigure}

    \vspace{0.5em} 

    \begin{subfigure}[t]{0.45\textwidth}
        \centering
        \includegraphics[width=\linewidth]{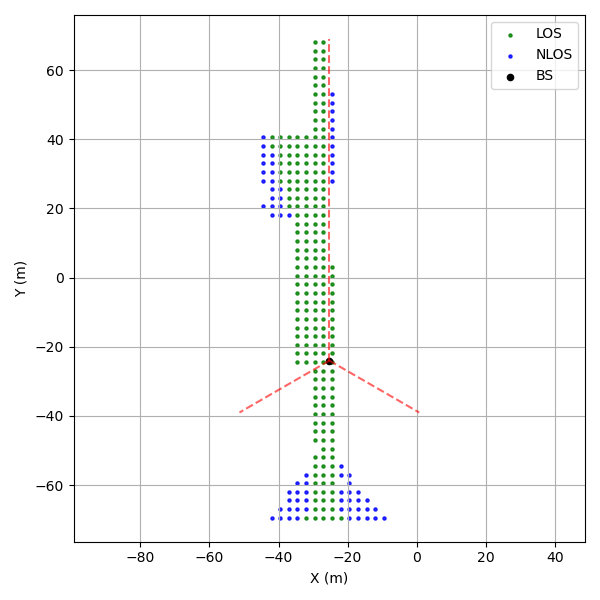}
        \caption{Chicago}
    \end{subfigure}
    \hfill
    \begin{subfigure}[t]{0.45\textwidth}
        \centering
        \includegraphics[width=\linewidth]{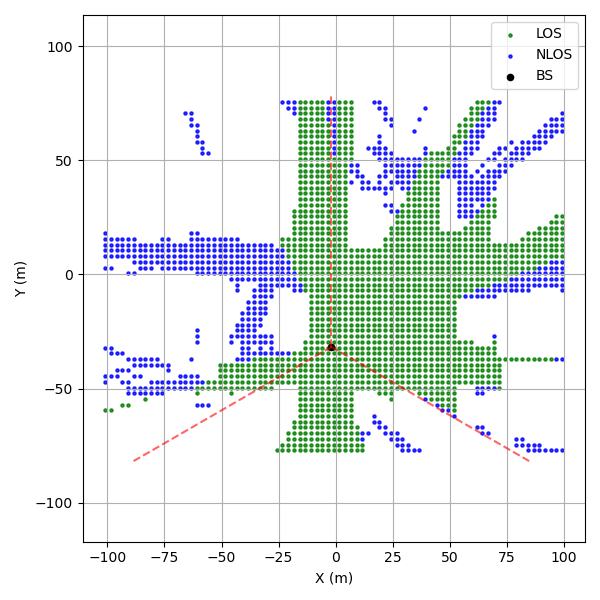}
        \caption{Houston}
    \end{subfigure}

    \caption{Base station and UE locations with LOS/NLOS distribution in different city scenarios. Green dots indicate LOS UEs, blue dots indicate NLOS UEs, and black dots represent base station positions.}
    \label{fig:urban-scenario-los-nlos}
\end{figure*}

\subsubsection{Loss function}

The total training objective consists of task-specific losses for positioning and beam prediction, as well as auxiliary losses from the feature fusion module.

For the positioning task, we adopt a mean squared error (MSE) loss between the predicted coordinate \( \hat{\mathbf{y}}_{\text{pos}} \) and the ground-truth position \( \mathbf{y}_{\text{pos}} \):
\begin{equation}
J_{\text{pos}} = \left\| \hat{\mathbf{y}}_{\text{pos}} - \mathbf{y}_{\text{pos}} \right\|^2.
\label{eq:pos_loss}
\end{equation}

For beam index classification, we use the cross-entropy loss between the predicted probability distribution \( \hat{\mathbf{y}}_{\text{bm}} \) and the one-hot encoded ground-truth label \( \mathbf{y}_{\text{bm}} \):
\begin{equation}
J_{\text{bm}} = - \sum_{j=1}^J \mathbf{y}_{\text{bm}}^{(j)} \log \hat{\mathbf{y}}_{\text{bm}}^{(j)},
\label{eq:bm_loss}
\end{equation}where $J$ is the total number of codebooks.

To further enhance the fusion quality between spatial and directional features, we incorporate two auxiliary losses from the fusion module: the adversarial loss \( J_{\text{adv}} \) defined in equation~\eqref{eq:adv_loss}, and the reconstruction loss \( J_{\text{Auto}} \) defined in equation~\eqref{eq:tr_loss}.

As a result, the total loss is then formulated as:
\begin{equation}
J_{\text{total}} = \lambda_{\text{pos}} J_{\text{pos}} + \lambda_{\text{bm}} J_{\text{bm}} + \lambda_{\text{adv}} J_{\text{adv}} + \lambda_{\text{Auto}} J_{\text{Auto}},
\label{eq:total_loss}
\end{equation}where \( \lambda_{\text{pos}} \), \( \lambda_{\text{bm}} \), \( \lambda_{\text{adv}} \), and \( \lambda_{\text{Auto}} \) are weighting coefficients for positioning, beam prediction, adversarial, and reconstruction losses, respectively. In our experiments, we set \( \lambda_{\text{pos}} = 0.01 \), \( \lambda_{\text{bm}} = 0.99 \), and tune \( \lambda_{\text{adv}} \) and \( \lambda_{\text{Auto}} \) empirically.

\subsubsection{Model Training}

The model is optimized using the AdamW optimizer with a maximum learning rate of $5 \times 10^{-4}$ and a weight decay of $1 \times 10^{-5}$. To stabilize the early training process, a linear warm-up strategy is adopted over the first 10 epochs. The total number of warm-up steps is calculated as $\text{warmup\_epochs} \times \text{steps\_per\_epoch}$, ensuring a smooth transition from small to large gradients.
\begin{table}[htbp]
\renewcommand{\arraystretch}{1.3}  
\centering
\caption{Parameter settings to generate dataset}
\begin{tabular}{lc}
\hline
\textbf{Parameter} & \textbf{Value} \\
\hline\hline
City 0 Scenario & New York, 1151 samples, LoS ratio 80.19\%\\
\hline
City 1 Scenario &Los Angeles, 690 samples, LoS ratio 51.45\% \\
\hline
City 2 Scenario & Chicago, 283 samples, LoS ratio 75.62\%\\
\hline
City 3 Scenario &  Houston, 2256 samples, LoS ratio 68.84\% \\
\hline
Carrier Frequency & 3.5 GHz \\
\hline
BS Height & 20 m \\
\hline
BS Antenna Array & $8 \times 8$ \\
\hline
BS Antenna Spacing & $0.5\lambda$ \\
\hline
Sector 1 & $-30^\circ$ to $90^\circ$ \\
\hline
Sector 2 & $90^\circ$ to $210^\circ$ \\
\hline
Sector 3 & $210^\circ$ to $330^\circ$ \\
\hline
BS Downtilt & $20^\circ$ \\
\hline
BS Antenna Spacing & $0.5\lambda$ \\
\hline
UE Height & 1.5 m \\
\hline
UE Antenna Array & $2 \times 2$ \\
\hline
UE Antenna Spacing & $0.5\lambda$ \\
\hline
Radiation Pattern & Isotropic (BS and UE) \\
\hline
OFDM Subcarriers & 256 \\
\hline
Selected Subcarriers & 0-255 \\
\hline
Total Bandwidth & 10 MHz \\
\hline
RX Filter & Disabled \\
\hline
$X_1 \times X_2$ & $4 \times 4$, $8 \times 8$, $16 \times 16$ \\
\hline
\end{tabular}
\label{tab:dataset-config}
\end{table}

\begin{figure*}[htbp]
    \centering

    \begin{subfigure}[t]{0.45\textwidth}
        \centering
        \includegraphics[width=\linewidth]{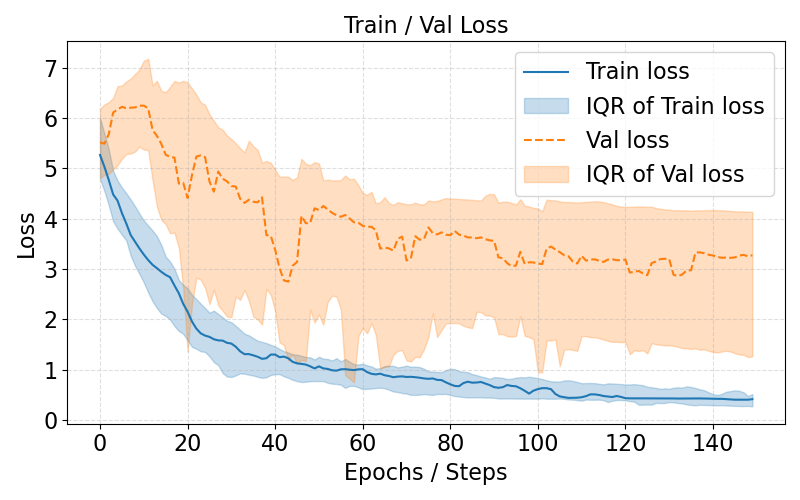}
        \caption{Proposed model~1 (AutoFusion)}
    \end{subfigure}
    \hfill
    \begin{subfigure}[t]{0.45\textwidth}
        \centering
        \includegraphics[width=\linewidth]{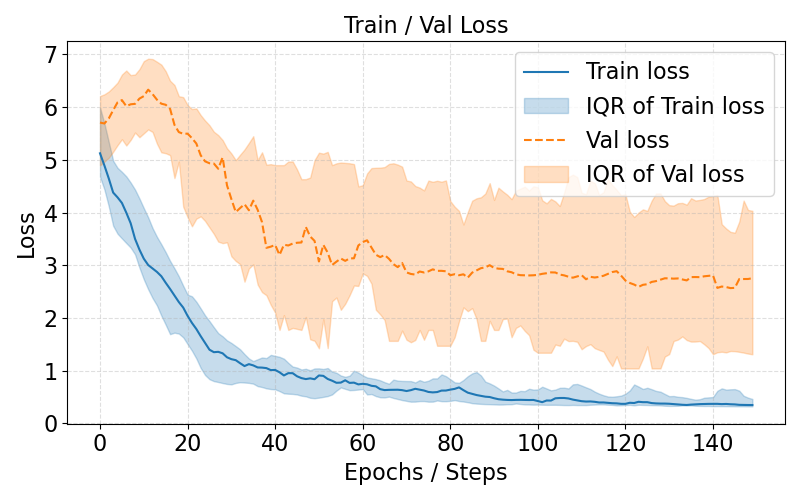}
        \caption{Proposed model~2 (GanFusion)}
    \end{subfigure}

    \caption{Training and validation loss convergence of the proposed models across different city scenarios. Solid lines denote the average loss, while shaded regions indicate the interquartile range (IQR).}
    \label{fig:convergence}
\end{figure*}

\begin{table*}[htbp]
\centering
\caption{BM accuracy (\%) for 16 $\rightarrow$ 64 beams under different SNRs (Top-1 Accuracy). \textbf{Bold} and \underline{underline} indicate the best and second best performance, respectively. ** (x) indicates that the beam prediction gain over the best baseline model is x. }
\setlength{\tabcolsep}{5.5pt}
\renewcommand{\arraystretch}{1.3}
\begin{adjustbox}{max width=\textwidth}
\begin{tabular}{cc*{7}{c}}
\toprule
{SNR} & {Scenario} & {HC}\cite{3gpp38.214,3gpp38.802} & {Baseline model 1} & {Baseline model 2}\cite{jalali2024fast} & {Baseline model 3}\cite{nguyen2022beam} & {Baseline model 4} & {Proposed model 1} & {Proposed model 2}  \\
\midrule
\multirow{4}{*}{0 dB}

& city0 &44.63 & \textbf{100.00} & 86.34 & 73.66 & \underline{99.51} & \textbf{100.00} (0.00)& \textbf{100.00} (0.00) \\
& city1 & 47.61 & \textbf{99.51} & 85.85 & 69.75 & \textbf{99.51} & \underline{99.27} (-0.24)& 99.01 (-0.50)\\
& city2 & 39.07 & 90.90 & 85.11 & 74.75 & 93.93 & \textbf{95.01} (1.08)& \underline{94.34} (0.41) \\
& city3 & 42.70 & 99.79 & 83.14 & 78.01 & \textbf{100.00} & \textbf{100.00} (0.00)& \underline{99.89} (-0.11)\\

\midrule
\multirow{4}{*}{20 dB}

& city0 & 66.71 & {98.05} & 76.58 & 58.53 & {97.23} & \textbf{100.00} (1.95)& \textbf{100.00} (1.95) \\
& city1 & 69.31 & {98.23} & {81.46} & 62.44 & \underline{98.84} & \textbf{100.00} (1.16)& \textbf{100.00} (1.16)\\
& city2 & 50.00 & {99.03} & 76.76 & 66.66 & 98.98 & \underline{98.99} (-0.04)& \textbf{100.00} (0.97)\\
& city3 & 50.78 & {99.10} & {76.36} & 52.11 & \underline{99.17} & \textbf{100.00} (0.83)& \textbf{100.00} (0.83) \\

\bottomrule
\end{tabular}
\end{adjustbox}
\label{tab:bm_64_fixed}
\end{table*}

\begin{table*}[htbp]
\centering
\caption{BM accuracy (\%) for 16 $\rightarrow$ 256 beams under different SNRs (Top-1 Accuracy). \textbf{Bold} and \underline{underline} indicate the best and second best performance, respectively. ** (x) indicates that the beam prediction gain over the best baseline model is x. }
\setlength{\tabcolsep}{5.5pt}
\renewcommand{\arraystretch}{1.3}
\begin{adjustbox}{max width=\textwidth}
\begin{tabular}{cc*{7}{c}}
\toprule
{SNR} & {Scenario} & {HC}\cite{3gpp38.214,3gpp38.802} & {Baseline model 1} & {Baseline model 2}\cite{jalali2024fast} & \textbf{Baseline model 3}\cite{nguyen2022beam} & {Baseline model 4} & {Proposed model 1} & {Proposed model 2}  \\
\midrule
\multirow{4}{*}{0 dB}
& city0 & 37.71 & 86.29 & 38.78 & 31.22 & 87.56 & \underline{89.07} (1.51)& \textbf{92.24} (4.68) \\
& city1 & 41.48 &87.32 & 40.49 & 32.68 & 82.43 & \textbf{92.73} (5.41)& \underline{91.29} (3.97)\\
& city2 & 34.82 & 65.65 & 43.43 & 35.35 & 66.66 & \underline{72.69} (6.03)& \textbf{77.76} (11.10) \\
& city3 & 34.51 & 89.54 & 36.48 & 29.49 & \underline{90.38} & \textbf{93.57} (3.18)& \textbf{93.57} (3.18)\\
\midrule
\multirow{4}{*}{20 dB}

& city0 & 55.90 & 87.99 & 57.31 & 46.58 &{90.97} & \textbf{93.73} (2.76)& \underline{92.26} (1.29)\\
& city1 & 63.82 & 89.36 & 58.05 & 42.92 & 87.80 & \textbf{94.22} (4.86)& \underline{92.27} (2.92)\\
& city2 & 36.44 & 67.74 & 56.56 & 41.41 & 71.00 & \textbf{75.78} (4.78)& \underline{78.74} (7.74) \\
& city3 & 44.14 & 90.64 & 55.29 & 49.33 & 90.95 & \textbf{96.01} (5.06)& \underline{94.46} (3.51)\\
\bottomrule
\end{tabular}
\end{adjustbox}
\label{tab:bm_256_fixed}
\end{table*}

After the warm-up phase, a learning rate scheduler based on the \texttt{ReduceLROnPlateau} mechanism is employed. The scheduler monitors validation performance in \textit{max} mode and reduces the learning rate by a factor of 0.8 if no relative improvement exceeding 1\% is observed for 10 consecutive epochs. A cooldown period of 2 epochs is introduced after each decay, and the learning rate is constrained to remain above a minimum threshold of $1 \times 10^{-8}$ to promote training stability and prevent overfitting.The model is trained for up to 150 epochs with a batch size of 256. Training is performed on a RTX 3090 GPU.

\section{Simulation Results}
\subsection{Dataset generation and simulation settings}
\subsubsection{Dataset generation} 
To meet the specifications of 3GPP TR 38.843~\cite{3gpp38843}, we utilize the DeepMIMO framework to generate our dataset. The specific parameters are summarized in Table.~\ref{tab:dataset-config}. Four outdoor city environments are selected as evaluation scenarios: New York (city 0), Los Angeles (city 1), Chicago (city 2), and Houston (city 3). All scenarios operate at a carrier frequency of 3.5 GHz with a three-sector configuration, featuring diverse propagation conditions and UE distributions. In each city scenario, a single BS is deployed, equipped with three $8 \times 8$ antenna arrays. Each array maintains half-wavelength spacing between elements and covers a $120^\circ$ sector with a $20^\circ$ downward tilt. This configuration ensures effective UE coverage while adhering to 3GPP standards.  

After generating CSI, data from the three sectors are combined to form complete $360^\circ$ coverage. Sector 1 covers angles from $-30^\circ$ to $90^\circ$, Sector 2 from $90^\circ$ to $210^\circ$, and Sector 3 from $210^\circ$ to $330^\circ$, with UEs outside each sector’s angular range filtered out. All CSI samples are annotated with either LOS or NLOS labels.

Fig. ~\ref{fig:urban-scenario-los-nlos} illustrates the base station and UE placements within these scenarios. City 0 contains 1151 samples, where UEs are densely clustered along street layouts, resulting in approximately 80.19\% LoS samples. In contrast, city 1 has 690 samples with a more balanced mix of LoS (51.45\%) and NLoS cases. City 2 comprises 283 samples, characterized by a highly directional UE distribution along narrow paths directly ahead of the base station, with 75.62\% of UEs in LoS. City 3 (Houston) includes 2256 samples, with an LoS ratio of 68.84\%, reflecting a more uniformly spread layout with coverage in all directions. Notably, city 2 represents an extreme case, with most UEs near the coverage edge, limiting spatial diversity. Conversely, city 3 offers a more typical and practical deployment scenario.

In our experiments, we configure three codebooks with angular resolutions $X_1 \times X_2$, corresponding to $4 \times 4$, $8 \times 8$, and $16 \times 16$, respectively.

\subsubsection{Baseline benchmarks}
To evaluate the performance of the proposed models, namely, proposed model~1 with AutoFusion and proposed model~2 with GanFusion, we compare them against a traditional hierarchical codebook–based beam training method and four representative baseline models.

\begin{itemize}
\item {HC}: A standard-compliant hierarchical beam training approach based on resolution-refined beam codebooks, which performs multi-stage coarse-to-fine beam sweeping and refinement and is widely adopted in current 5G NR BM procedures~\cite{3gpp38.214,3gpp38.802}.
\item {Baseline 1}: A ResNet-based model\cite{he2016deep} equipped with a three-class classification head trained solely for beam prediction.
\item {Baseline 2}: A model based on the structure proposed in \cite{jalali2024fast}, which combines CNNs and MLPs to predict the optimal beam.
\item {Baseline 3}: An LSTM-based framework introduced in \cite{nguyen2022beam}, utilizing LSTM followed by MLP layers for beam prediction.
\item {Baseline 4}: A RegNet-based model with a three-class classification head dedicated to beam prediction.
\end{itemize}

\subsection{Results analysis}
In this section, we evaluate the Top-1 beam prediction accuracy of the proposed and baseline models across scenarios in cities 0–3, under various SNR levels. 
Specifically, we first analyze the model convergence behaviors, followed by accuracy comparisons, and finally examine the resulting communication performance.
\subsubsection{Convergence analysis}
We first analyze the training convergence behaviors of the proposed models based on the training and validation loss curves shown in Fig.~\ref{fig:convergence}. 
Both proposed model~1 and proposed model~2 exhibit stable and smooth convergence across different city scenarios. 
The training loss decreases monotonically, while the validation loss gradually stabilizes with limited fluctuation, indicating effective optimization and good generalization without severe overfitting.

\subsubsection{In-distribution (ID) performance}
We first compare the ID performance of baseline models 1–4 and proposed models 1 and 2 in scenarios within cities 0-3. All models are trained on a subset of one dataset and tested on a disjoint subset from the same dataset, ensuring in-distribution evaluation.

Table \ref{tab:bm_64_fixed} presents the results for 16 $\rightarrow$ 64 beam prediction, where the goal is to use measurements from a coarse 16-beam codebook to predict the optimal beam within a finer 64-beam codebook. It shows that baseline models 1 (ResNet-based) and 4 (RegNet-based), which rely solely on beam features, achieve performance comparable to the proposed models that incorporate positioning information via the proposed feature fusion module. This suggests that, for 16 $\rightarrow$ 64, positional data contributes little to beam prediction accuracy. Conversely, baseline models 2 (CNN-based) and 3 (LSTM-based) perform worse, likely due to less effective feature extraction. All models perform better in higher SNR scenarios, with baseline models 2 and 3 exhibiting significant accuracy improvements as SNR increases from 0 to 20 dB. However, models 1 and 4, along with the proposed models, already approach nearly 100\% accuracy at lower SNRs, resulting in minimal gains from additional SNR.

Further comparison for 16 $\rightarrow$ 256 beam prediction is shown in Table \ref{tab:bm_256_fixed}, where measurements from a coarse 16-beam codebook are used to predict the optimal beam within a much finer 256-beam codebook. As expected, increasing the number of candidate beams makes the task more complex, leading to notable performance degradation. The accuracy of proposed models 1 and 2 drops by approximately 5–10\% across cities 1–4, at both 0 and 20 dB SNR. Baseline models 1 and 4 experience over 10\% accuracy decline, while baseline models 2 and 3 exhibit the most significant drop of around 30\%. Notably, the proposed feature fusion modules offer greater advantages in low SNR conditions, with performance gains of 1.51–11.10\% over the best baseline in 0 dB scenarios, diminishing to 1.29–7.74\% at 20 dB. These gains are primarily attributed to the effective integration of position-related features. An additional observation is that beam prediction in scenario city 2 remains more challenging than in the other scenarios, with all models showing poorer results at both SNR levels.

In conclusion, these findings highlight the importance of combining positional and beam information through effective fusion mechanisms. Our proposed dual-fusion models consistently outperform baselines across diverse environments and SNR conditions.

\begin{table*}[htbp]
\centering
\caption{Zero-shot generalization performance of Top-1 (16 $\rightarrow$ 256 beams) beam prediction accuracy (\%): model trained using dataset of one city scenario and tested in the remaining 3 city scenarios. \textbf{Bold} and \underline{underline} indicate the best and second best performance, respectively. ** (x) indicates that the beam prediction gain over the best baseline model is x. }
\setlength{\tabcolsep}{4.5pt}
\renewcommand{\arraystretch}{1.3}
\begin{adjustbox}{max width=\textwidth}
\begin{tabular}{cccccccc}
\toprule
 {Training dataset} & {Testing dataset}  & {Baseline model 1} & {Baseline model 2} \cite{jalali2024fast} & {Baseline model 3}\cite{nguyen2022beam} & {Baseline model 4} & {Proposed model 1} & {Proposed model 2}  \\ 
\midrule
\multirow{6}{*}{City 0}
& City 1 (0dB)& 78.88 & 34.63 & 28.78 & 76.10 & \textbf{83.80} (4.92)& \underline{82.39} (3.51) \\
& City 1 (20dB)& 80.95 & 46.83 & 40.49 & 77.15 & \textbf{84.41} (3.46)& \underline{84.93} (3.98)\\
& City 2 (0dB)& 60.63 & 28.28 & 26.26 & {63.66} &  \underline{65.64} (1.98)& \textbf{67.68} (4.02) \\
& City 2 (20dB)& 61.62 & 31.31 & 32.32 & 64.55 & \underline{66.88} (2.33)& \textbf{68.62} (4.07)\\
& City 3 (0dB)& 81.81 & 31.76 & 28.16 &77.08  & \textbf{85.76} (3.95)& \underline{84.73} (2.92) \\

& City 3 (20dB)& 82.84 & 44.19 & 33.20 & 84.07 & \underline{87.86} (3.79)& \textbf{88.23} (4.16)   \\
\midrule
\multirow{6}{*}{City 1}
& City 0 (0dB)& 75.56 & 31.22 & 29.51 & 75.51 & \underline{76.76} (1.2) &\textbf{79.51} (3.95) \\
& City 0 (20dB)& 76.90 & 45.12 & 36.10 & {76.24}  & \underline{77.32} (0.43)& \textbf{80.88} (3.98) \\
& City 2 (0dB)& 55.56 & 31.31 & 13.13 & 61.62 & \underline{66.77} (5.15)& \textbf{66.86 } (5.24)\\
& City 2 (20dB)& 60.61 & 32.32 & 29.29 & {66.67} & \textbf{71.67} (5.0)& \underline{69.65} (2.98) \\
& City 3 (0dB)& 71.84 & 30.63 & 28.16 & {78.93} & \textbf{84.29} (5.36)& \underline{83.11} (4.18) \\
& City 3 (20dB)& 72.66 & 39.16 & 30.94 & 80.68 & \textbf{84.12} (3.44)& \underline{83.24} (2.56) \\
\midrule
\multirow{6}{*}{City 2}
& City 0 (0dB)& 72.44 & 35.12 & 29.02 & 67.56 & \textbf{81.27} (8.83)& \underline{77.37} (4.93) \\
& City 0 (20dB)& 78.68 & 45.12 & 35.12 & 76.49 & \textbf{84.15} (5.47)& \underline{82.34} (3.66) \\
& City 1 (0dB)& 72.68 & 33.17 & 28.29 & 68.29 & \underline{80.00} (7.32)& \textbf{82.44} (9.76) \\
& City 1 (20dB)& 75.57 & 58.05 & 38.54 & 74.63 & \underline{82.05} (5.42)& \textbf{84.39} (7.76)\\
& City 3 (0dB)& 67.42 & 30.63 & 27.75 & 65.67 & \underline{76.07} (8.62)& \textbf{77.00} (9.38) \\
& City 3 (20dB)& 77.80 & 39.16 & 29.25 & 77.47 & \textbf{81.57} (3.77)& \underline{80.02} (3.22)\\
\midrule
\multirow{6}{*}{City 3}
& City 0 (0dB)& 80.22 & 34.15 & 27.80 & 79.12 & \underline{83.49} (3.27) & \textbf{84.39} (4.17) \\
& City 0 (20dB)& 83.17 & 44.39 & 43.66 & 84.88 & \textbf{87.80} (3.92)& \underline{86.15} (1.27) \\
& City 1 (0dB)& 78.71 & 35.61 & 24.39 & 76.59 & \textbf{82.44} (3.73) & \underline{81.95} (3.24)  \\
& City 1 (20dB)& 82.51 & 46.34 & 43.90 & 80.49 & \textbf{85.46} (2.95)& \underline{84.15} (1.64) \\
& City 2 (0dB) & 60.64 & 39.39 & 22.22 & 60.71 & \textbf{67.58} (6.87) & \underline{64.55} (3.84) \\
& City 2 (20dB)& 65.64 & 30.30 & 27.27 & 61.62 & \textbf{69.62} (3.98)& \underline{67.65} (2.01) \\
\bottomrule
\end{tabular}
\end{adjustbox}
\label{tab:gen_city0}
\end{table*}

\subsubsection{Out-of-distribution (OoD) Generalization Performance}

To assess the generalization capabilities of our models, we evaluated their performance across different city scenarios. The previous section highlighted the complexity of transitioning from 16 to 256 beam prediction. Therefore, we conducted robustness tests under this challenging scenario, focusing on Top-1 beam prediction accuracy.

Each model was trained in one city and tested across others, maintaining consistent SNR conditions. Our results, detailed in Table \ref{tab:gen_city0}, reveal that our models consistently outperform baseline models across various city scenarios and SNR levels. Notably, the performance improvement in 0 dB scenarios is more pronounced than in 20 dB. For instance, when models trained in city 0 were tested in cities 1-3, the gain ranged from 1.98-4.92\% at 0 dB and 2.23-3.98\% at 20 dB. The models trained in city 2 and tested in city 0, 1 and 3 scenarios, the performance gain in 0 and 20 dB are 4.93-9.76\% and 3.22-7.76\%, respectively. Similar gains were observed when training on other cities, underscoring superior generalization in lower SNR conditions.

Aligned with the ID performance, the poorest OoD results were in city 2, with accuracy of 64.55-67.68\% at 0 dB and 67.65-69.65\% at 20 dB, versus over 76\% in other cities. This discrepancy likely stems from city 2's unique layout. 

These results demonstrate that our fusion-based beam prediction framework not only excels in the original environments but also generalizes well across diverse urban deployments. The consistently strong performance confirms the effectiveness of leveraging positioning information via feature fusion, validating the design of our positioning-assisted beam prediction strategy.

\begin{table}[t]
\centering
\captionsetup{width=0.9\columnwidth}
\caption{Average communication rate (bps) across 16$\rightarrow$64 and 16$\rightarrow$256 beam tasks under different SNRs}
\begin{tabular}{c|l|cccc}
\hline
{SNR} & {Method} & {city0} & {city1} & {city2} & {city3} \\
\hline
\multirow{5}{*}{0 dB}
& HC\cite{3gpp38.214,3gpp38.802}                & 0.62 & 0.62 & 0.58 & 0.61 \\
& Baseline model 2 \cite{jalali2024fast}  & 0.72 & 0.69 & 0.69 & 0.73 \\
& Baseline model 3\cite{nguyen2022beam}  & 0.63 & 0.64 & 0.60 & 0.63 \\
& Proposed model 1  & 0.92 & 0.92 & 0.92 & 0.90 \\
& Proposed model 2  & 0.93 & 0.91 & 0.93 & 0.92 \\
\hline
\multirow{5}{*}{20 dB}
& HC\cite{3gpp38.214,3gpp38.802}                & 6.11 & 6.10 & 5.58 & 5.76 \\
& Baseline model 2 \cite{jalali2024fast}  & 6.18 & 6.11 & 5.91 & 6.02 \\
& Baseline model 3\cite{nguyen2022beam}  & 6.08 & 6.04 & 5.66 & 5.85 \\
& Proposed model 1  & 6.53 & 6.51 & 6.52 & 6.52 \\
& Proposed model 2  & 6.55 & 6.53 & 6.49 & 6.52 \\
\hline
\end{tabular}
\label{tab:comm_rate_avg}
\end{table}

\begin{table}[h]
\centering
\caption{Per-sample inference time and FLOPs comparison}
\begin{tabular}{l|c|c}
\hline
{Method} & {Inference Time (ms)} & {FLOPs (GFLOPs)} \\
\hline
Baseline model 1  & 0.29 & 0.33 \\
Baseline model 2\cite{jalali2024fast}  & 0.14 & 0.02 \\
Baseline model 3\cite{nguyen2022beam}  & 0.22 & 0.03 \\
Baseline model 4  & 0.44 & 0.33 \\
Proposed model 1  & 0.47 & 0.66 \\
Proposed model 2  & 0.65 & 0.67 \\
\hline
\end{tabular}
\label{tab:inference_time}
\end{table}

\subsubsection{Ablation Study on Backbone and Positional Feature Fusion}

To further quantify the contribution of the backbone architecture and the proposed feature fusion strategy, we conduct a systematic ablation study using the results reported in Tables~\ref{tab:bm_64_fixed} and~\ref{tab:bm_256_fixed}. The ablation is structured along three progressively enhanced model groups:
(1) baseline model~1 using a ResNet-50 backbone,
(2) baseline model~4 using a RegNet backbone, and
(3) the proposed models integrating positional information through the feature fusion module.
This setup allows us to isolate the effects of (i) backbone improvement and (ii) auxiliary positional-information fusion.
In addition, a traditional hierarchical codebook (HC)–based beam training method is included as a non-learning baseline for reference.

For the 16 $\rightarrow$ 64 task, all strong backbones achieve near-saturated Top-1 accuracy across SNR conditions, with RegNet slightly outperforming ResNet-50. However, the proposed fusion-based models offer only marginal gains (typically within $\pm 1\%$), indicating that positional information brings limited benefit when the prediction granularity is moderate and beam-domain features are already highly discriminative.

In contrast, the 16 $\rightarrow$ 256 task presents a substantially more challenging setting, where the benefits of architectural and feature-level enhancements become evident. RegNet consistently outperforms ResNet-50, with improvements of 2--5\% at 0~dB and 1--3\% at 20~dB, confirming its superior feature extraction capabilities. More importantly, the proposed fusion-based models yield significant accuracy gains over both baselines. Across city scenarios, the improvement over RegNet ranges from 1.5--7.7\% at 0~dB and 2.7--7.9\% at 20~dB, demonstrating the value of incorporating positional cues when the beam search space is large and ambiguous.By contrast, the traditional HC–based approach generally underperforms learning-based methods with strong backbones. 
Nevertheless, HC achieves competitive or even superior performance compared to baseline models~2 and~3 in some scenarios, indicating that well-designed hierarchical beam search can still be effective under certain conditions.

Overall, this ablation study highlights two key insights. First, RegNet serves as a stronger backbone than ResNet-50 for beam prediction, especially in low-SNR and large-codebook settings. Second, integrating positional features via the proposed feature fusion module provides substantial performance gains in fine-grained beam prediction tasks, validating the design choice of leveraging auxiliary spatial information to enhance robustness and accuracy.

\subsubsection{Communication performance}
The resulting communication performance is summarized in Table~\ref{tab:comm_rate_avg}, where the average achievable rate is evaluated by aggregating the 16$\rightarrow$64 and 16$\rightarrow$256 beam tasks. The achievable rate is computed based on the effective beamformed channel, given by $\log_2\!\left(1 + \frac{\|\mathbf{H}\mathbf{f}_i\|^2}{\mathcal{N}}\right)$.For clarity and consistency with prior studies, we report communication-rate results for representative baselines only. 

At 0~dB, the proposed models consistently achieve higher communication rates than all baseline methods across all city scenarios. 
Specifically, proposed model~2 attains the highest average rate, followed closely by proposed model~1, indicating that more accurate beam selection directly translates into improved link capacity under low-SNR conditions.

At 20~dB, although the performance gap among different methods becomes narrower due to the saturation of the Shannon capacity, the proposed models still maintain a clear advantage over the baselines. 
Both proposed model~1 and proposed model~2 achieve rates around 6.5~bps across all city scenarios, outperforming baseline models~2 and~3 as well as the HC–based method. 
These results demonstrate that the benefits of learning-based beam prediction persist even in high-SNR regimes, leading to more efficient utilization of the available spatial degrees of freedom.

Overall, the communication-rate results are well aligned with the beam prediction accuracy trends, confirming that the proposed models not only improve beam selection performance but also deliver tangible gains in end-to-end communication efficiency.

\subsubsection{Inference Time}
We also measured the inference times of our models compared to baseline alternatives. As shown in Table~\ref{tab:inference_time}, baseline models~2 and~3 achieve the fastest inference with 0.14~ms and 0.22~ms per sample, respectively. For reference, these two lightweight CNN- and RNN-based models also have relatively low computational complexity, with FLOPs in the range of 0.02--0.04~GFLOPs. Baseline models~1 and~4, both using large CNN backbones (ResNet and RegNet), require approximately 0.33~GFLOPs each.

Our proposed models incorporate a dual-branch backbone design and additional fusion modules, resulting in higher computational cost. Proposed~1 reaches about 0.66~GFLOPs, while Proposed~2 slightly increases to 0.67~GFLOPs due to the added GanFusion module. The increased complexity mainly stems from the dual-branch RegNet backbone, which provides stronger feature extraction capability compared to single-branch designs. 
In addition, the introduction of the feature fusion module slightly increases the computational cost, particularly for Proposed~2, where the GanFusion module introduces additional operations.
Despite the increased complexity, their inference times of 0.47~ms and 0.65~ms remain well within the 1~ms processing budget defined in the 5G~NR specification~\cite{3gpp38211}, ensuring real-time feasibility.

Given that sub-millisecond variations rarely cause significant channel fluctuation, the additional computation does not compromise practical deployment. 
Importantly, the modest increase in inference time is offset by noticeable gains in beam prediction accuracy in both ID and OoD scenarios, confirming the effectiveness of our positioning-assisted fusion framework.

\section{Conclusion}
In this paper, we proposed a positioning-assisted deep learning framework for beam prediction that improves prediction accuracy while reducing beam training overhead. By leveraging positioning-aware supervision through coordinate regression, the model learns effective spatial representations, which are further fused with beam-domain features via adaptive and adversarial fusion modules. Built on a dual-branch RegNet architecture, the proposed framework consistently outperforms strong baselines across different urban scenarios and frequency bands. Extensive evaluations on DeepMIMO datasets demonstrate improved generalization, especially under challenging conditions such as high-resolution beam codebooks and low-SNR scenarios.

As future work, we will extend the framework to distributed learning paradigms, such as federated learning, to address the decentralized nature and data scarcity of wireless networks, and investigate cross-frequency and multi-band extensions to enhance robustness in heterogeneous deployments.

\bibliographystyle{IEEEtran}
\bibliography{references} 

\end{document}